\documentclass[prl,aps,floats,twocolumn,showpacs]{revtex4}

\usepackage{graphicx}
\usepackage{dcolumn}
\usepackage{bm}
\usepackage{amsmath}

\begin{document}

\title{Divergent Thermal Conductivity in Three-dimensional Nonlinear Lattices }

\author{Hayato Shiba}
\email{shiba@acolyte.t.u-tokyo.ac.jp},
\author{Satoshi Yukawa}
\altaffiliation{Present address: Department of Earth and Space Science, Graduate School of Science, Osaka University, Osaka 560-0043}  
\author{Nobuyasu Ito}
\email{ito@ap.t.u-tokyo.ac.jp}
\affiliation{Department of Applied Physics, University of Tokyo, \\ Hongo, Bunkyo-ku, Tokyo 113-8656}

\date{\today}

\begin{abstract}
Heat conduction in three-dimensional nonlinear lattices is investigated 
using a particle dynamics simulation.
The system is a simple three-dimensional extension of 
the Fermi-Pasta-Ulam $\beta$ (FPU-$\beta$) nonlinear lattices, 
in which the interparticle potential has a biquadratic term together with
a harmonic term. The system size is $L\times L\times 2L$, 
and the heat is made to flow in the $2L$ direction with using 
the Nos\'e-Hoover method.  
Although a linear temperature profile is realized, 
the ratio of energy flux to temperature gradient shows logarithmic 
divergence with $L$. The autocorrelation function of energy flux 
$C(t)$ is observed to show power-law decay as $t^{-0.98\pm 0.25 }$, 
which is slower than the decay in conventional momentum-conserving 
three-dimensional systems ($t^{-3/2}$). Similar behavior is also 
observed in the four-dimensional system.
\end{abstract}

\maketitle

Thermal conduction has been one of the main issues of statistical mechanics for more than a century. Thermal conduction is usually accurately described by the Fourier law
\begin{equation}
\bm{J} = -\kappa\nabla T,
\end{equation}
where $\bm{J}$ is the heat flux, $\kappa$ is the heat conductivity, 
and $T$ is the local temperature. The transport coefficient 
is described by the Green-Kubo formula\cite{ Kubo, KuboNakajima, Green} 
\begin{equation}
\kappa = \frac{k_B\beta^2}{V}\int_0^\infty  C(t)\mbox{d}t, 
\label{eq:GK}
\end{equation}
where $C(t)=\langle\bm{J}(t)\cdot\bm{J}(0)\rangle $ denotes the equilibrium autocorrelation function of $\bm{J}(t)$.

In our understanding of the conventional long-time tails of 
the autocorrelation function, $C(t)\sim t^{-d/2}$ in a $d$-dimensional 
system\cite{Alder, Pomeau, Ernst},
the integral in eq. (\ref{eq:GK}) is expected to 
diverge in one- and two-dimensional 
systems, and to converge in three- and higher-dimensional systems. 
For a finite system with size $L$, the size dependence of the effective
transport coefficient $\kappa (L)$, which is defined as the ratio
of energy flux to temperature gradient, is obtained by replacing
the upper limit of the integral in eq. (\ref{eq:GK}) with the time
range $L/v_s$ ($v_s$ denotes a typical phonon velocity). Together
with the long-time tail behavior, we obtain $\kappa (L) = a + b L^{1-d/2}$,
where $a$ and $b$ are constants depending on the system, and $L^{1-d/2}$
should be interpreted to be a logarithmically behaving function $\ln L$
for $d=2$. This argument predicts $\kappa (L) \sim L^{0.5}$, $\ln L$
and $a+b/L^{0.5}$ for $d=1$, $2$ and $3$, respectively. 
Therefore, $\kappa (L)$ is expected to diverge for $d=1,2$ 
and to converge for $d\geq 3$\cite{Shimada,LepriRep}. 

Such size dependence has been verified in two- and three-dimensional
fluid systems. In the two-dimensional hard-disk system, logarithmic 
divergence has been confirmed. In the three-dimensional case, 
$1/\sqrt{L}$ convergence has been confirmed in 
the hard-spheres system\cite{Shimada,Murakami} and in 
the Lennard-Jones system\cite{Ogushi,OgushiB}.

In this article, we consider nonlinear lattice systems with 
total momentum conservation as a model system of insulated solids.
In one-dimensional nonlinear lattices\cite{Machida,Hatano, Shimada,LepriBeta, LepriEur, LepriRep, Narayan},
the power-law divergence of $\kappa (L)$ has been confirmed, 
where the estimated value of the exponent is about $0.37$. 
This value is smaller than the one expected from the above argument. 
In the two-dimensional case, 
logarithmic divergence has been observed\cite{LippiLivi,LepriChaos}. 

The behavior of three-dimensional nonlinear lattice models 
has not been clarified, however. 
Only a Fermi-Pasta-Ulam-$\beta$ (FPU-$\beta$)-like three-dimensional 
model has been studied, and $1/\sqrt{L}$ convergence was 
observed\cite{Shimada}. This model has a natural length 
in the potential function, and a free boundary condition was used 
in the direction of heat flow. 
As a result, the system does not have a crystalline 
structure in the steady state.

The purpose of this study is to determine whether three-dimensional
nonlinear lattices with momentum conserving interaction
show $1/\sqrt{L}$ convergence.

Our model is described by the Hamiltonian
\begin{equation}
\mathcal{H} = \sum_{i=1}^N \frac{\bm{p}_i^2}{2}
+\sum_{\langle i,j\rangle}\left[ \frac{1}{2}|\bm{r}_i-\bm{r}_j|^2 
+\frac{g}{4}|\bm{r}_i-\bm{r}_j|^4\right], \label{eq:Hamiltonian}
\end{equation}
It is a simple extension of the FPU-$\beta$\cite{FPU} chain to higher 
dimensional lattices. Here, $\bm{p}_i$ and $\bm{r}_i$ denote
the momentum and displacement of a particle on lattice point $i$ 
respectively, and are three-dimensional vectors.
All the particles have the unit mass.
The summation over $\langle i,j\rangle$ denotes 
the nearest-neighbor lattice points.

A cubic lattice with a size of $L_x\times L_y\times L_z$ is considered.
A periodic boundary condition is used in the $L_x$ and $L_y$ directions.
A fixed boundary condition is used in the $L_z$ direction; that is,
the particles at both ends in the $L_z$ direction are coupled to
rigid walls through the same interaction potential as that between 
nearest-neighbor particles. Furthermore, the particles at both ends in the $L_z$
direction undergo temperature control by the Nos\'e-Hoover method\cite{NoseP}.
The temperature at one end is denoted by $T_L$ and the other end by $T_R$; 
therefore, energy flows from the $T_L$ end to the $T_R$ end 
along the $L_z$ direction after the system reaches a steady state. 
In summary, the equations of motion are

\begin{subequations}
\begin{eqnarray}
\dot{\bm{r}}_i &=& \bm{p}_i \label{eq:eqm1} \\
\dot{\bm{p}}_i &=& \left\{\begin{array}{ll}{}
 -\displaystyle\frac{\partial\mathcal{H}}{\partial\bm{r}_i} & \textrm{(in the bulk)} \\  
 -\displaystyle\frac{\partial\mathcal{H}}{\partial\bm{r}_i} -\zeta_i\bm{p}_i & \textrm{(at both ends)}, \label{eq:eqm2}\\ 
\end{array}\right. 
\end{eqnarray}
\end{subequations}
where $\zeta_i$ denotes the Nos\'e-Hoover thermostat variables, which obey
\begin{equation}
\dot{\zeta}_i = \frac{1}{Q}\left( \frac{\bm{p}_i^2}{3k_BT}-1\right). \label{eq:HB}
\end{equation}
\begin{figure}[b]
\centerline{\resizebox{0.5\textwidth}{!}{\includegraphics{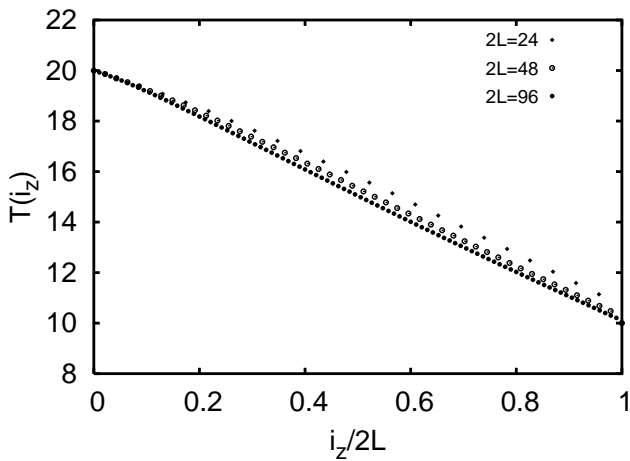}}}
\caption{Temperature profile of three-dimensional FPU-$\beta$ lattice. The size of the system is taken as $L\times L\times 2L$. Each sequence represents the result for different sizes $2L = 24, 48, \textrm{and }96$ from top to bottom. The horizontal axis shows the $z$-coordinate rescaled by the system size $2L$, and the vertical axis shows the local temperature averaged over the cross-sectional cut in the $xy$-plane. The $3\sigma$ error is within the marks for each point.}
\label{f1}
\end{figure}
\begin{figure}[t]
\centerline{\resizebox{0.5\textwidth}{!}{\includegraphics{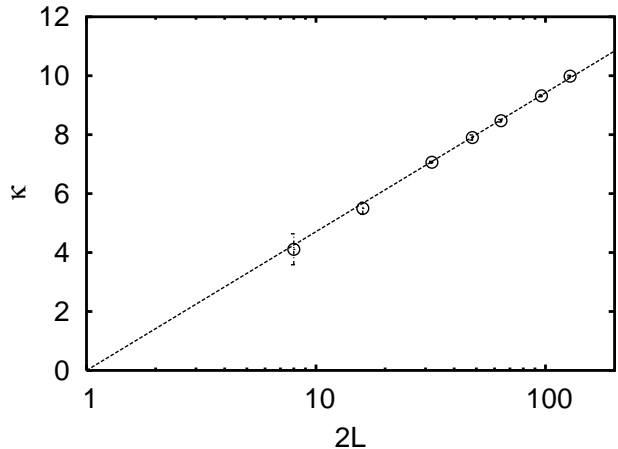}}}
\caption{Size dependence of thermal conductivity in three-dimensional FPU-$\beta$ lattices with system size of $L\times L\times 2L$ given in log scale. Logarithmic divergence is observed, as shown by the fitted line, ($2.047\pm 0.004 )\times\log 2L$.}
\label{f2}
\end{figure}

$T$ denotes the target temperatures, $T_L$ and $T_R$, and $Q$ denotes
the coupling parameter between $\zeta_i$ and $\bm{p}_i$. 
Boltzmann's constant $k_B$ is taken as 1.0. In the following, we fix the parameters as
\begin{equation}
T_L = 20.0,\ T_R=10.0,\ \textrm{and}\  Q=1.0.
\end{equation}
In this temperature region, the nonlinear term in the Hamiltonian in eq. (\ref{eq:Hamiltonian}) has the same order of contribution as the linear term. Thus, the temperature is sufficiently high for the dynamical evolution to reproduce the thermal state.

First, we study the system of size $L\times L\times 2L$ to avoid the effects of anisotropy and dimensional crossover. Simulations are carried out using system sizes from $L=4$ to 64.

Simulations start from a state with all displacements $\bm{r}_i=0$ 
and with randomly selected momenta $\bm{p}_i$, so that 
the local kinetic energy profile satisfies a linear temperature 
profile from $T_L$ to $T_R$. From this 
initial state, the system finally reaches a steady state.
This initial relaxation process takes about $t =5\times 10^4$ for $L=64$.
Afterwards, we sample the local temperature $T(i_z)$, where the particles 
are labeled in order as $(i_x, i_y, i_z)$. This is given by the average of 
the local temperature $T(i)$ of each particle $i$ in the sectional plane of 
$z=i_z$, with $T(i)$ given by the long-time average of the kinetic energy:
\begin{equation}
T(i) = \langle\bm{p}_i^2/3\rangle.
\end{equation}
The temperature profile is shown in Fig. \ref{f1}. 
A typical simulation time is about $1.0\times 10^5$ for $L=64$.h
Therefore, the total number of simulation steps per particle
for one sample is about $1.0\times 10^7$ which takes slightly 
more than 1 month using a single core of a 2.2 GHz Opteron processor
for the $L=64$ system. Five to eight samples are accumulated for the
results. The temperature profile $T(z)$ becomes linear, 
which shows that thermalization is sufficient in this state. 
\begin{figure}[tb]
\centerline{\resizebox{0.50\textwidth}{!}{\includegraphics{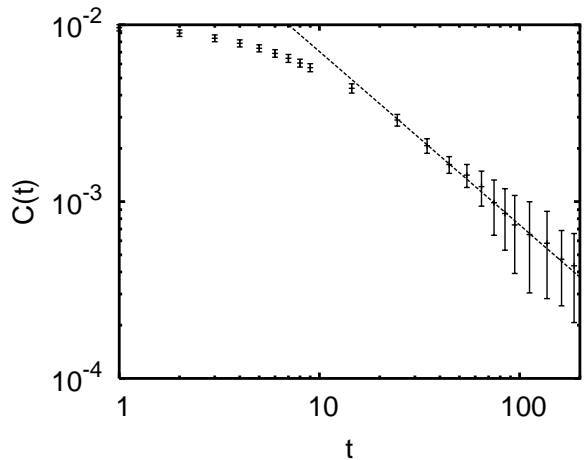}}}
\caption{Autocorrelation function $C(t) =\langle J_z(t)J_z(0)\rangle$ in three-dimensional FPU-$\beta$-lattice with system size of $32\times 32\times 64$. The temperature is set to $T=15$.  The periodic boundary condition is applied, and the power-law exponent of the long-time tail behavior is shown by the fitted line ($C(t)\sim t^{-0.98\pm 0.25}$)}.
\label{f3}
\end{figure}

We estimate the local heat flux at site $i$ with the energy flow using 
\begin{equation}
\quad j_i = -\frac{1}{2}(\dot{\bm{r}}_{i+1_z} +\dot{\bm{r}}_i) \cdot\frac{\partial}{\partial \bm{r}_i}V(\bm{r}_{i+1_z}-\bm{r}_i),
\end{equation}
where $i+1_z$ denotes the nearest-neighbor site of the $i$th 
particle in the $z$-direction. Heat flux per particle, $J_z$, is estimated using
\begin{equation}
J_z = \left\langle \sum_i \frac{j_i}{N}\right\rangle,
\end{equation}
where $\langle\cdot\rangle$ denotes the time average after 
the simulation reaches to a steady state and $N$ denotes
the total number of particles. Then, the thermal conductivity 
$\kappa (L)$ is estimated using
\begin{equation}
\kappa (L) = J_z \Big/ \left(\frac{dT}{dz}\right).
\end{equation}
Figure \ref{f2} shows the estimated values of this $\kappa (L)$.
A logarithmic divergence is clearly observed for the systems up to 
$64\times 64\times 128$.
Divergence of $\kappa (L)$ with $L$ means that the system does not have 
finite conductivity at the macroscopic limit.

This logarithmic divergence of $\kappa (L)$ is consistent with the 
long-time behavior of the autocorrelation function, 
$C(t) =\langle J_z(t)J_z(0)\rangle$ in the equilibrium state.
Figure \ref{f3} shows the estimated values of this $C(t)$ obtained by
microcanonical simulation of the same nonlinear lattice without 
temperature control. The system size is $32\times 32\times 64$, 
and the total energy in the system is adjusted to the internal 
energy expectation value at temperature $T=15$.
A periodic boundary condition is used in all directions in this simulation.
In Fig. 3, it is observed that $C(t)$ decays asymptotically as 
$1/t^{0.98\pm 0.25}$ in the long-time limit, which suggests the 
logarithmic divergence of $\kappa (L)$, as we observed in our nonequilibrium 
simulation. This is evidence that the divergence of $\kappa (L)$ is simply 
due to the bulk property of the system, and it is not caused by a 
boundary effect or by the temperature control.

\begin{figure}[t]
\centerline{\resizebox{0.50\textwidth}{!}{\includegraphics{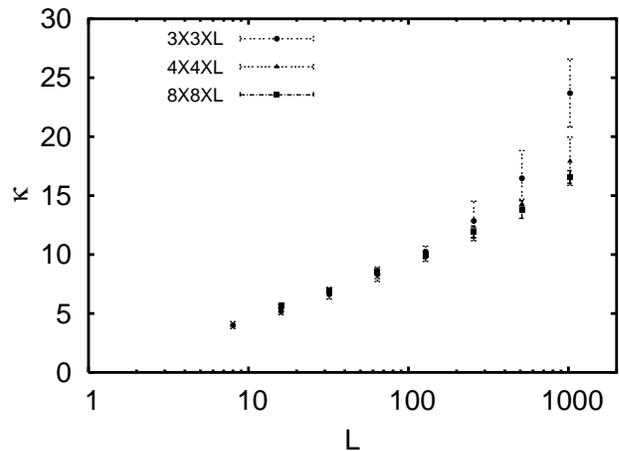}}}
\caption{Size dependence of thermal conductivity in quasi-one-dimensional FPU-$\beta$ lattices. Circle, triangle, and square plots respectively represent the thermal conductivities for system sizes of $3\times 3\times L,\ 4\times 4\times L$, and $8\times 8\times L$. 
}
\label{f4}
\end{figure}

Secondly, the thermal conductivity of quasi-one-dimensional systems 
is studied. For fixed $L_x$ and $L_y$, the $L_z$ dependence of the 
thermal conductivity, $\kappa (L_z)$, is estimated. Details of the 
computer simulation and the model parameters are the same as above. 
The systems with $L_x=L_y=$3, 4, and 8 are simulated for $L_z=8$ to 
1024. The results are shown in Fig. \ref{f4}.
Estimated values of thermal conductivity are consistent with the 
bulk values of the system with the same $L_z$ as shown in Fig. 2, 
for up to $L_z\sim 128$, and the longer systems have larger $\kappa$ 
and show power-law divergence, which was confirmed in various 
one-dimensional systems. The crossover length from the logarithmic 
divergence to the power-law increases with thickness ($L_x=L_y$). 
The length is about 256 or 512, even for the thinnest $L_x=L_y=3$ system. 
Therefore, the crossover length is about 100 times the cross-section 
size. This result may explain the reason why the convergence of
thermal conductivity was confirmed in previous studies that used 
quasi-one-dimensional systems.

\begin{figure}[b]
\centerline{\resizebox{0.50\textwidth}{!}{\includegraphics{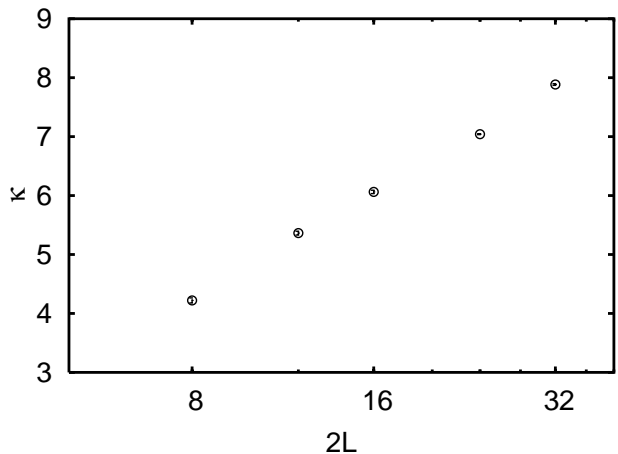}}}
\caption{Size dependence of thermal conductivity in four-dimensional FPU $\beta$ lattices using semilog scale. The size of the system is $L\times L\times L\times 2L$. }
\label{f5}
\end{figure}

So far, we have observed that conductivity is logarithmically divergent not 
only in two-dimensional, but also in three-dimensional FPU-$\beta$ lattices. 
We show here that the conductivity of the four-dimensional FPU-$\beta$ 
lattice also has the tendency of logarithmic divergence. We prepare the 
hypercubic lattice with a system size of $L\times L\times L\times 2L$ in 
four-dimensional space, 
and simulate the system using eqs. (\ref{eq:Hamiltonian}), (\ref{eq:eqm1}), 
and (\ref{eq:eqm2}), and boundary conditions similar to those used in 
the three-dimensional model. The same parameters are used as in the 
three-dimensional case except for the temperatures, which are set to 
be $T_L=15.0$ and $T_R=7.5$. The calculated thermal conductivity is 
shown in Fig. \ref{f5}. It is notable that the tendency of divergence 
still appears, as in the three-dimensional case, although we could only 
calculate up to the system size of $2L\sim 32$ because larger 
systems require too much computational load. The result seems to indicate
that no dimensionality effect exists in the behavior of the conductivity at
the isotropic thermodynamic limit when the dimension is higher than 2.

The behavior of the model with a natural length and fixed boundary conditions
has not yet been investigated, although the convergence of $\kappa (L)$
has been observed in the free-boundary-applied FPU-$\beta$ lattice with a 
natural length\cite{Shimada}. We simulated such systems with a size of 
$5\times 5\times L$. We set the natural length as $l_0=100.0$, 
and the boundary condition in the $x$ and $y$ directions as periodic,
and in $z$ direction, the particles at both ends are linked to missing atom 
which cannot move. The missing atoms on the left and the right sides are 
separated by a distance of $(L+1)\times l_0$. In such a situation, 
the conductivity $\kappa(L)$ is confirmed to show diverging behavior 
for a system size of up to $L=512$.

In this study, we investigated heat transport in nonlinear lattices 
with momentum conservation as a model of insulated solids. The model 
we used contained no impurities or randomness in mass or interaction, 
and only the effects of nonlinear interaction and dimensionality 
contribute to thermalization. We find that such nonlinear lattice 
systems do not have finite thermal conductivity at the thermodynamic 
limit, even if they are three- and four-dimensional systems. Contrary to 
the standard understanding of $t^{-d/2}$ long-time tail behavior, $t^{-1}$ 
behavior is widely observed not only in two-dimensional lattices, but also 
in three- and four-dimensional lattices, and may also appear in 
higher-dimensional lattices. Therefore, the standard long-time tail argument 
cannot be applied in the case of nonlinear lattices. It is natural, because
the conventional long-time tail is considered a consequence of
viscous modes, which do not exist in our model solids. Our next challenge
is to clarify the kinematical origin of the $t^{-1}$ behavior. 
Of course, studying systems with larger sizes is a topic of future study.

We have not investigated in detail how the conductivity behaves when 
we increase the values of nonlinear strength $g$ and temperature $T$. 
However, we confirm a similar divergence of $\kappa (L)$ when $g$ is 
up to 1.0 and $T$ is up to approximately 100.0. Detailed investigations on 
the effect of stronger nonlinearity are problems to be investigated.

It is clear that there is still a long way to go before we can model 
normal heat conduction using nonlinear lattice models. Or, our results may be 
an indication of a new mechanism of heat flow. The problems 
concerning the minimum conditions for normal heat conduction
and the behavior of heat flow in insulated solids remain unsolved. 
Heat conduction in three-dimensional models requires
further investigation to realize a conclusive model 
for heat conduction in crystalline insulated solids.

\acknowledgements
The authors thank the Supercomputer Center, Institute for Solid State Physics, 
University of Tokyo for the use of the HITACHI SR11000. This work is partially 
supported by the Japan Society for the Promotion of Science (No. 14080204). 
The authors gratefully thank Keiji Saito, Takashi Shimada and Koji Azuma for valuable discussions, 
and Synge Todo for helpful advice.

\end{document}